# Assessing the Brain Wave Hypothesis:

# Call for Commentary

Robert Worden

Theoretical Neurobiology Group, University College London, London, United Kingdom

rpworden@me.com

Draft 1.1; July 2024

Abstract:

It has been proposed that there is a wave excitation in animal brains, whose role is to represent three-dimensional local space in a working memory. Evidence for the wave comes from the mammalian thalamus, the central body of the insect brain, and from computational models of spatial cognition. This is described in related papers.

I assess the Bayesian probability that the wave exists, from this evidence. The probability of the wave in the brain is robustly greater than 0.4. If there is such a wave, we may need to re-think our whole understanding of the brain, in a break from classical neuroscience. I ask other researchers to comment on the wave hypothesis and on this assessment. In a companion paper, I outline possible ways to test it.

## 1. Introduction

It has been proposed [1,2,3] that there is a wave excitation in the brain, which acts as a working memory. representing the three-dimensional space around any animal. This is a departure from the classical neural model of the brain, and it differs from other proposals of electromagnetic fields in the brain [5,6] – which have concerned computation, rather than memory.

If this wave hypothesis is correct, it will require a far-reaching revision of neuroscience. It is important to know whether it has some chance of being correct (say, with probability greater than 10%); and if it has, to confirm it or refute it.

The reasons for proposing the wave hypothesis are described in [1,2,3], and some can be seen on YouTube at https://www.youtube.com/watch?v=gMPa_xoOctU. This paper does not repeat the evidence, but uses it to estimate the probability that there is such a wave in the brain.

A Bayesian method is used to combine different pieces of evidence about the wave hypothesis into a single estimate of its probability. This is done in tables, which refer to the detailed evidence. The resulting probability is robustly large – being greater than 0.4 by two different weighting methods. It would take strong evidence against the wave to reduce its probability to less than 0.1, and I know of no such evidence. Equally, there is not enough evidence to believe the hypothesis – as for instance in a minimal one-standard deviation (P = 0.95) criterion for belief.

If the wave exists, it has far-reaching implications for neuroscience. It would change the paradigm for brain research, much as Rutherford's nuclear model of the atom changed the paradigm of atomic physics. There would be a strong case [4] to identify the wave as the source of consciousness. This would advance our understanding of one of the greatest scientific challenges.

At present the physical nature of the wave is completely unknown. It is like dark matter in the brain.

Because of the large issues which turn on the existence of the wave, the present level of uncertainty about it is unsatisfactory. It is a high priority to resolve it.

The weakness of the wave hypothesis is that it rests on the work of one person. As far as I know, only one person has pursued this idea, explored evidence about it, and assessed the evidence in this paper. The dependence on one researcher urgently needs to be remedied. As a first step, I ask you to provide commentary on the wave hypothesis – both generally, and from your own specialist knowledge. If you wish, your commentary will be visible at www.bayeslanguage.org/bb/commentary.html.

Professor Karl Friston has endorsed this call for commentary, in section 7 of this paper.

## 2. The Wave Hypothesis of Spatial Cognition

I first describe the hypothesis which is to be assessed.

Animals need to understand the 3-D space around them, just to move their limbs. This is a primitive and essential requirement for a brain.

In Bayesian theories of cognition, animals build an internal 3-D model of local space, using Bayesian inference from their sense data (notably vision). The wave hypothesis is a hypothesis about how this 3-D model is stored in the brain. It says that the model is stored not in neural firing rates, but in a wave.

It proposes that there is some approximately round region of the brain, of diameter D, which holds a wave excitation. Neurons couple to the wave as transmitters and receivers; the wave acts as a working memory. Waves in the region can have any wavelength larger than some minimum wavelength $\lambda_{min}$, and can be in any direction[1]. A wave with wavelength $\lambda$ represents some object in space at a distance $c/\lambda$, (where c is a constant) and in the direction of travel of the wave (i.e. perpendicular to the wave fronts).

This round region of the brain, and two waves in it, is shown in figure 1. The figure also shows the positions in space of the two objects represented by the two waves. Shorter wavelengths represent larger distances.

---

[1] That is why the region in the brain needs to be approximately spherical – to allow waves in any direction.

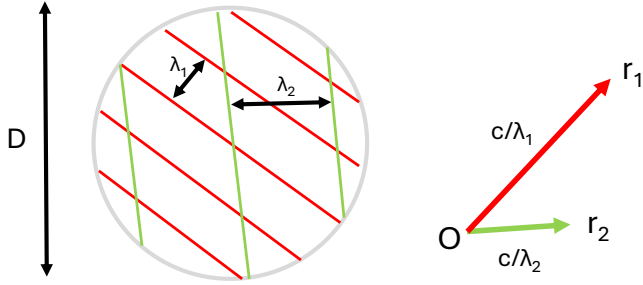

*Figure 1: a spherical region in the brain, of diameter D, holding two waves with wavelength $\lambda_1$ (red) and $\lambda_2$ (green). These represent objects in space at positions $r_1$ and $r_2$.*

The wave representation of space is one of the simplest ways to represent positions of things in 3-D space – simpler than, say, using coordinates in some 3-D coordinate system, represented by neural firing rates. It has two important properties:

1. As the wave can hold many independent waves with different wavelengths and directions, it can store the positions of many different things – of the order of $(D/\lambda_{min})^3$ things, which can be a very large number.
2. Each location is stored with a precision approximately $(\lambda_{min}/D)$. With realistic values for $\lambda_{min}$ and D, this precision can easily be better than 1%. It is very hard to get that degree of precision from neural firing rates.

The relationship between the wave and the space it represents is a Fourier transform – a relation used in many branches of science. The wave represents space in the same way as a hologram does.

That is the wave hypothesis to be tested.

## 3. Two Ways to Estimate the Probability of the Wave Hypothesis

In the papers that describe the evidence for the wave [1,2,3] several types of evidence are described. How can the different evidence be combined, to estimate an overall probability that the wave exists?

The way to combine different pieces of evidence about some hypothesis is to use Bayes' theorem to compute the posterior probability of the hypothesis, as modified by each piece of evidence, starting from some prior probability [7]. The prior probability is an initial view of how unlikely the hypothesis is, before taking account of the evidence, based on the complexity of the hypothesis and other prior considerations. For the Bayesian analysis other assumptions may need to be made, such as assuming the independence of different pieces of evidence, or weighting different evidence.

I shall make this Bayesian analysis with two different sets of weightings.

The hypothesis to be tested is the following:

> **Hypothesis**: There is a three-dimensional wave excitation in every animal brain (located in the thalamus in mammals, or in the central body of the insect brain), whose purpose is to act as a spatial working memory, holding a 3-D model of the animal's surroundings, for periods as long as fractions of a second.

This hypothesis has the virtue of specificity, making it potentially easy to refute. It is a binary hypothesis; either there really is such a wave, or there is not. It differs from other proposals about waves in the brain such as [5,6].

If there is no wave, our model of the brain reverts to the classical neural model of the brain, where all computation is done by neural synaptic connections. If there is a wave, our model of spatial cognition (and all cognition that depends on it) becomes a 'nuclear' model, with the wave at the logical and physical centre of the brain, much like the nucleus in Rutherford's atom. Neurons depend on the wave, maintain it and use it, as electrons orbit round the nucleus, or as planets orbit round the sun.

The first approximation in the Bayesian calculation is to treat the different pieces of evidence as independent. To the extent that they are not independent, the result should be adjusted; but adjustments are not made in this paper. We then give weights to the different pieces of evidence.

Because the hypothesis is binary, each piece of evidence is treated as a small set of binary decisions – like a small number of coin tosses, with the result of each coin toss being either for the wave, or against it. By adjusting the number of coin tosses for each piece of evidence, the weight of that evidence is defined. By adjusting the outcomes of the coin tosses between 'for' and 'against', the contribution of each piece of evidence (for or against the hypothesis) is defined.

For instance, if some piece of evidence is characterized as five binary decisions, all in favour of the wave, that evidence is weighted as being quite strong, and makes a positive contribution. If it is taken as five binary decisions, but they are split 3:2, it has the same weight of 5; but its contribution is less. Either the weight or the contribution may change with further work.

This Bayesian analysis is mathematically equivalent to estimating the bias of a coin, from several independent tests of tossing the coin (tossing from the left hand, using a randomised tossing machine, tossing from an aeroplane, etc.). The weight of each test is its number of coin tosses. The best estimate of the bias of the coin (the probability that it will come out heads on the next toss) is just the sum of all tosses which come out heads, divided by the total number of tosses across all tests. This combines the different tests according to their weights.

There is a subjective element to this calculation, which comes from giving a weight and a contribution to each piece of evidence. I have made these assessments visible in the tables which follow, and in an Excel spreadsheet, which you can download from www.bayeslanguage.org/bb/assesswave.zip . You can use the spreadsheet to flex any of the weights or contributions, or to add other pieces of evidence, to see the impact on the resulting probability.

Because of the element of subjectivity in assigning a weight and a contribution to each piece of evidence, I have estimated the probability of the wave hypothesis in a second, simpler way. This is just to assign one 'vote' to each piece of evidence, making each piece of evidence entirely for or against the wave hypothesis, with equal weight. In this estimate, the probability of the hypothesis is just the sum of all 'for' votes, divided by the total number of votes (the total number of pieces of evidence considered).

Weights are not entirely subjective. If some piece of evidence has weight W, then on its own (without any other evidence), it has the potential to imply a probability as small as $1/(W+2)$, or as large as $(W+1/W+2)$, or any of $(W - 1)$ probabilities uniformly spaced in between them, depending on its contribution (e.g. you can try these formulae for W = 1, one binary choice; the resulting probability is 1/3 or 2/3). This can be used to assess the suitability of any weight assignment. I have used no weights greater than 4 for any positive evidence for the wave, and have only used integer weights.

Because it uses less refined weights, the second estimate is expected to be less reliable than the first. Comparing the two estimates gives an idea of how much the result is sensitive to the weight assessments.

## 4. Estimates of the Probability of the Wave

I first show the overall results – the two estimates of the probability of the wave hypothesis. The estimates are shown in the summary table:

| Evidence | for | against | weight | Votes total | votes for |
|---|---|---|---|---|---|
| (1) prior considerations | 0 | 25 | 25 | 3 | 0 |
| (2) mammalian thalamus | 14 | 2 | 16 | 5 | 5 |
| (3) insect central body | 9 | 1 | 10 | 4 | 4 |
| (4) computational models | 12 | 2 | 14 | 6 | 6 |
| (5) biophysical mechanisms | 6 | 14 | 20 | 7 | 3 |
| (6) phenomenal consciousness | 4 | 0 | 4 | 4 | 4 |
| (7) other considerations | 2 | 0 | 2 | 2 | 2 |
| Grand Total | 47 | 44 | 91 | 31 | 24 |

31 pieces of evidence are considered, in the seven groups shown. The sum of weights from all groups is 91. The sum of contributions in favour of the wave hypothesis is 47, making the first estimate of the probability of the wave hypothesis (47/91) = 52%.

A figure as low as 52% is only obtained by giving a large weight of 25 to negative prior 'evidence' against the hypothesis. This is mainly from Occam's Razor – an unwillingness to make extra assumptions about the brain. This strong negative weight has been chosen to show that the resulting estimate is insensitive to prior assumptions.

The table shows how the estimate is supported. The two groups (1) and (5) of negative evidence (prior considerations, and biophysical mechanisms) are more than balanced by groups (2) – (4) of positive evidence for the wave – from the mammalian thalamus, the insect central body, and computational models. The other groups (6) and (7) make smaller contributions.

The second estimate is to assign an equal weight of 1.0 to all 'votes' of all the thirty pieces of evidence. On this basis, the probability of the wave hypothesis is 24/31 = 77%.

Both estimates are well above the threshold where the hypothesis needs to be taken seriously, but are short of a minimal threshold of 95% (equivalent to one standard deviation in a measurement) at which we might start to believe it.

I next show the assessments of the 31 pieces of evidence, in their groups. The description of the evidence in each group is brief. For details, see papers [1,2,3]. The weightings and contributions of each piece of evidence not justified individually; you may assess them from the detailed descriptions.

All the tables are in an Excel spreadsheet accompanying this paper, or downloadable. You are invited to modify the spreadsheet, altering the weighting or contribution of any piece of evidence, or adding new pieces of evidence, to make your own estimate of the probability of the wave hypothesis.

Tables for each group of evidence follow.

| (1) prior considerations | for | against | weight | Votes total | votes for |
|---|---|---|---|---|---|
| (a) Occam's Razor: a wave in the brain is an extra hypothesis; it is better to do without it | 0 | 20 | 20 | 1 | 0 |
| (b) the classical neural model of cognition is sufficient to describe the brain. | 0 | 2 | 2 | 1 | 0 |
| (c) we have never observed a long-lived wave in the brain | 0 | 3 | 3 | 1 | 0 |
| Total | 0 | 25 | 25 | 3 | 0 |

**Prior Considerations**: I include three prior considerations: (a) a general "Occam's Razor" reluctance to admit any new hypothesis; (b) a view that the classical neural model of the brain is sufficient; and (c) the fact that the proposed wave has not yet been observed. Point (a) of Occam's Razor has been given large weight, to show that the final result is not sensitive to it. For reasons described in [3], (b) and (c) are given little weight.

| (2) mammalian thalamus | for | against | weight | Votes total | votes for |
|---|---|---|---|---|---|
| (a) the thalamus is approximately round in all mammal species, so it is well suited to hold a wave. Other parts of the brain are irregular and are not suited for a wave. | 2 | 0 | 2 | 1 | 1 |
| (b) the thalamus has the right neural connections to do spatial sense data integration | 2 | 1 | 3 | 1 | 1 |
| (c) the thalamus is the gateway to cortex for sense data of all modalities, except olfaction. Smell is of little use for the fast precise location of objects. | 2 | 1 | 3 | 1 | 1 |
| (d) unless you assume a wave, the anatomy of the thalamus (nuclei collected together, with few or no connections between them) does not make sense in energy terms. Thalamic nuclei need to be clustered together to be immersed in the same wave. | 4 | 0 | 4 | 1 | 1 |
| (e) unless you assume a wave, the anatomy of the TRN (a thin sheet round the dorsal thalamus) does not make sense in energy terms. In the wave hypothesis, a thin sheet may serve some purpose. | 4 | 0 | 4 | 1 | 1 |
| Total | 14 | 2 | 16 | 5 | 5 |

**Mammalian Thalamus**: This is one of three main groups of evidence for the wave. The thalamus has just the right neural connections needed to store a model of local 3-D space; and its distinctive round shape is well suited to hold a wave [1].

From an energy viewpoint, the anatomy of the thalamus does not make sense in a purely neural model of cognition; but it may make sense if there is a wave. This is a smoking gun for the wave hypothesis [15 - 17]. It is the only positive evidence in any table given a weight as large as 4; all other positive weights are 3 or less.

| (3) insect central body | for | against | weight | Votes total | votes for |
|---|---|---|---|---|---|
| (a) the central body of the insect brain is approximately round in all insect species, so it is well suited to hold a wave. Other parts of the insect brain are irregular. | 2 | 0 | 2 | 1 | 1 |
| (b) the insect central body has the right neural connections to do spatial sense data integration | 2 | 1 | 3 | 1 | 1 |
| (c) the shape of the central body is remarkably conserved across all insect species. This shape must serve some purpose. | 3 | 0 | 3 | 1 | 1 |
| (d) there are not enough neurons in a typical insect brain for a good representation of 3-D space; but insects do spatial cognition well | 2 | 0 | 2 | 1 | 1 |
| Total | 9 | 1 | 10 | 4 | 4 |

**Insect central body**: This is the second main evidence for the wave [1]. The insect central body, like the thalamus, has just the right neural connections to hold a model of local 3-D space; and its distinctive round shape is well suited to hold a wave. In the central body, there is not the same 'smoking gun' clue for a wave as there is in the thalamus, as the insect central body does not have discrete nuclei.

However, there is other important evidence for a wave in insects. The shape of the insect central body is remarkably conserved across all insect species – unlike most other parts of the insect brain. This conserved shape must serve a purpose; it cannot be understood if the role of the central body is only to connect neurons by synapses. Holding the wave is a possible purpose, which requires it to have a constant shape.

Finally, the insect brain has far fewer neurons available for neural storage of locations, so there is a stronger case for non-neural storage of 3-D positions.

Insects and mammals are at opposite ends of the animal evolutionary tree, having diverged approximately 500 million years ago. They suggest that the wave in the brain exists across the whole evolutionary tree from their joint ancestor – that is, in nearly all animal brains.

| (4) computational models | for | against | weight | Votes total | votes for |
|---|---|---|---|---|---|
| (a) there are no working neural computational models of 3-D spatial cognition; existing models ignore neural noise and errors, and focus on learning, rather than building a 3-D map of local space | 2 | 1 | 3 | 1 | 1 |
| (b) the errors from neural representation of spatial coordinates in sub-second intervals are too large to hold a good model of 3-D space, or to compute | 2 | 1 | 3 | 1 | 1 |
| (c) the classical McCulloch-Pitts- Hebb model of neurons under-utilises the capabilities of the eukaryotic cell; neurons are more capable than that. | 1 | 0 | 1 | 1 | 1 |
| (d) a wave representation of 3-D space is simple, direct and natural; it simplifies the required computations. | 2 | 0 | 2 | 1 | 1 |
| (e) a wave representation of spatial coordinates has high capacity, and can give errors less than 1%. | 3 | 0 | 3 | 1 | 1 |
| (f) a wave representation of space can give fast response times | 2 | 0 | 2 | 1 | 1 |
| Total | 12 | 2 | 14 | 6 | 6 |

| (5) biophysical mechanisms | for | against | weight | Votes total | votes for |
|---|---|---|---|---|---|
| (a) there is no proposed physical mechanism for the wave | 0 | 10 | 10 | 1 | 0 |
| (b) there is increasing evidence for quantum coherent effects in biological matter and in the brain | 1 | 0 | 1 | 1 | 1 |
| (c) It is hard to devise a coherent quantum effect that can store information for fractions of a second | 0 | 1 | 1 | 1 | 0 |
| (d) We cannot rule out a Bose-Einstein condensate in the brain, which could store information for long times | 1 | 0 | 1 | 1 | 1 |
| (e) The many qualia in our conscious experience would imply that the wave has many degrees of freedom, if the wave is the source of consciousness | 1 | 2 | 3 | 1 | 0 |
| (f) there is no way proposed for neurons to couple to the wave; but neurons can couple to many excitations at very low intensities. | 1 | 1 | 2 | 1 | 0 |
| (g) Evolution has discovered many mechanisms beyond those we have invented. | 2 | 0 | 2 | 1 | 1 |
| Total | 6 | 14 | 20 | 7 | 3 |

**Computational Models of the brain**: This evidence for the wave is the failure of classical neuroscience to build working neural models of spatial cognition. Nearly all animals do basic spatial cognition well, and it is essential for their survival – being needed at every moment of the day. But neural computational models of 3-D spatial cognition have made little progress [1, 3].

Existing neural models [8 - 14] focus on learning shapes, rather than the more basic and necessary task of building a 3-D map of local space. Crucially, they model the outputs of neurons, including spatial position estimates, as high-precision real variables, rather than as noisy spike trains. This side-steps a core problem of spatial cognition – that neuron firing is too slow and too imprecise to represent 3-D space within sub-second response times [1,3]. A wave may be able to do that.

To re-state this challenge: classical neural models of the brain cannot do spatial cognition. Neurons can do the required computations, but they cannot store the information with the required speed and precision. This challenge can only be met by building a working neural computational model of spatial cognition. I have not yet seen one.

If you have spent time building neural models of the brain, the suggestion that they cannot do a core cognitive task (building a 3-D model of local space) may not be welcome. If you do not like the idea, and do not like the related proposal of a wave in the brain, please do not ignore the messenger. Please send commentary on this paper, saying what neural path you see as the way forward.

**Biophysical mechanisms**: Because the wave is required to serve as a working memory for periods up to fractions of a second, it is probably not simply electromagnetic [3], so it has not yet been observed in the brain. This empirical evidence has been included already in 'prior considerations'. The leading theoretical evidence against a wave in the brain is that no biophysical mechanism for it has yet been proposed. I have given this negative evidence the large weight of 10 – in spite of the fact that evolution is more ingenious than we are, and has found many powerful mechanisms before we have invented them.

There are other lines of evidence, described in [3], pulling in both directions. The net effect is negative, reflecting that the question of the biophysical nature of the wave is wide open – an unsolved problem.

| (5) phenomenal consciousness | for | against | weight | Votes total | votes for |
|---|---|---|---|---|---|
| (a) If there is a wave in the brain representing local 3-D space, there is a strong case to identify the wave as the source of consciousness | 1 | 0 | 1 | 1 | 1 |
| (b) The spatial properties of the wave are a good fit to the spatial properties of our conscious experience ; it fits the main evidence we have about consciousness | 1 | 0 | 1 | 1 | 1 |
| (c) A wave theory of consciousness solves three serious problems facing neural theories of consciousness (selection of conscious neurons, decoding neural representations of space, neural imprecision) | 1 | 0 | 1 | 1 | 1 |
| (d) The form of our spatial consciousness shows that the human brain has a very good internal model of 3-D space - better than neurons are capable of, but feasible for a wave. | 1 | 0 | 1 | 1 | 1 |
| Total | 4 | 0 | 4 | 4 | 4 |

**Phenomenal Consciousness**: If there is a wave in the brain representing 3-D local space, then the role of the wave is a good fit to the form of our spatial conscious awareness – which consists of a 3-D model of the space around us. This makes a case to identify the wave as the source of consciousness [4]. It fits the data about spatial consciousness in a simple way, and solves some serious problems which have beset neural theories of consciousness for many years. So the form of consciousness gives indirect support for the wave hypothesis.

The wave hypothesis of consciousness is discussed at https://www.youtube.com/live/IvY4acL8YeA .

Because theories of consciousness are controversial, I have given this evidence small weight, so that the overall result does not depend on it.

| (7) other considerations | for | against | weight | Votes total | votes for |
|---|---|---|---|---|---|
| (a) all vertebrate species have a thalamus, which may be a blackboard for spatial cognition | 1 | 0 | 1 | 1 | 1 |
| (b) some single-celled animals show cognition which may be evidence for non-neural cognition. | 1 | 0 | 1 | 1 | 1 |
| Total | 2 | 0 | 2 | 2 | 2 |

**Other considerations**: There are other considerations, but they currently have little weight compared to the evidence in groups (2) – (4); it is a matter for future research [3] to develop them. I have included only two pieces of evidence.

## 5. Robustness of the Result

Under both weighting schemes, the wave hypothesis has probability greater than 40% – enough to imply that it should be taken seriously. Perhaps the most remarkable aspect of the result is its robustness.

'Robust' means that you can alter the weights of different evidence, or add other evidence, across quite a wide range, and still the core result does not change.

One measure of robustness is that the result emerges from both of two very different weighting schemes for the evidence – one of which attempts to be as realistic as possible, the other as simple as possible.

The result does not depend on using large weights. The only weights greater than 4 which have been used are weights of negative arguments, against the wave hypothesis; and the positive weight 4 has only been used for the 'smoking gun' evidence from thalamic neuroanatomy. Therefore no piece of positive evidence has been taken on its own to imply a probability greater than 5/6; I have deliberately not applied strong weights to any positive evidence for the wave. Negative arguments are included in the overall assessment, with large weights 20 (for Occam's Razor) and 10 (for biophysical mechanisms). A lot of further negative evidence with large weights would be needed to make the overall probability less than 10%.

It is the accumulation of small pieces of evidence for the wave, each inconclusive on its own, which implies that the probability of a wave is robustly large. The evidence literally does add up.

You are invited to use the spreadsheet to test the result for yourself – to flex the weights and contributions of evidence, or to add new evidence, and to explore the results. I would be interested to hear your results.

## 6. Implications, if the Wave Exists

If the hypothesis is confirmed, of a wave acting as a working memory for 3-D spatial information, then its implications for neuroscience will be profound.

Spatial cognition is of primary importance in the animal brain. At every moment of the day, an animal needs to understand the three-dimensional locations of the things around it, just to move – the most basic thing that animals do, which distinguishes them from plants. Their survival depends on spatial cognition, and there has been huge sustained selection pressure to do it as well as possible. Animals devote most of their brains to spatial cognition; they spend large resources gathering sense data for it; and they appear to do it very well.

Spatial cognition is the core of what brains do. Yet from many years' work in neuroscience, there is little understanding at a neural computational level (Marr's [20]

Level 3) how it works. The classical neural model of cognition has not succeeded in this regard. Neural computational models of spatial cognition have only rarely been built [8 - 14]; they have focused on learning of shapes, rather than core spatial cognition (building a 3-D spatial map of reality); and they have not addressed an overarching problem, that a neural representation of 3-D space by spike trains is slow and imprecise – too slow and imprecise to represent spatial positions. Neural models have not modelled spike trains.

If the wave hypothesis is confirmed, it will radically change this situation. It opens the way to building working computational models of spatial cognition – the core function of an animal brain – as classical neural models have so far failed to do. This will re-cast our understanding of the main function of the brain, and it will have implications for nearly all aspects of brain function.

It would completely revise the core model of the brain – from seeing the brain as a rather inchoate mass of connected neurons ('the most complex machine in the universe') to a focused nuclear model – of a central wave with a clearly defined function (to represent 3-D space), and neurons which have evolved around it to create, maintain and use the model.

There is a further important consequence. If there is a wave in the brain, there is a compelling case to identify the wave as the source of consciousness [4]. The function of the wave is a good fit to the spatial nature of consciousness, and it solves serious problems which have beset neural models of consciousness for many years. The conditional probability

P (Wave is the source of consciousness|Wave exists)

is large, close to 1.0.

Therefore if there is a wave, it may resolve[2] one of the hardest problems in science. This on its own calls for investigation of the wave hypothesis.

## 7. Call for Commentary

A cardinal principle of science is reproducibility. A scientific result is not really a result if it remains the work of only one person, or even one group; it needs to reach a consensus amongst researchers, or at least a debate. Other researchers should be able to reproduce the reasoning, or the experiments, for themselves.

The weakness of the wave hypothesis is that it currently rests on the work of one person. As far as I know, only one person has pursued this idea, explored evidence about it, and assessed the evidence in this paper. (while other waves in the brain have been proposed [5,6], their function has been computation rather than working memory)

In view of the potential importance of the hypothesis, this dependence on one person needs to be remedied. As a first step, I ask you to provide commentary on the wave hypothesis (however brief), both generally, and from your specialist knowledge. You may want to comment on this paper or any of the supporting papers.

The spreadsheet is an invitation to challenge the conclusions of the paper, by introducing new evidence or by altering weights of evidence. Download the spreadsheet from www.bayeslanguage.org/bb/assesswave.zip .

You may also comment on the broader issue – do neural models of the brain need some major new ingredient? Or is it enough to carry on developing the current models? Is there an appetite for change in neuroscience?

Your comments are valued particularly if you are unconvinced by the wave hypothesis. Please email commentary to rpworden@me.com . If you wish, your comments will be posted at www.bayeslanguage.org/bb/commentary.html.

Professor Karl Friston has endorsed this call for commentary on the wave hypothesis. Referring to this description of the hypothesis:

*In Bayesian theories of cognition, animals build an internal 3-D model of local space, using Bayesian inference from their sense data (notably vision). The wave hypothesis is a hypothesis about how this 3-D model is stored in the brain. It says that the model is stored not in neural firing rates, but in a wave.*

Karl Friston writes:

This, in a nutshell, is the (Bayesian) "waves in the brain" hypothesis. A hypothesis that challenges conventional (i.e., neuron doctrine) views of belief updating in the Bayesian brain. And offers a perspective on neuromorphic computation that is supported not only by theoretical accounts of functional brain architectures — but is receiving recent empirical evidence that speaks to ephaptic mechanisms. **It would be very useful to see other people's take on this proposal for mortal computation in the brain**.

(Emphasis mine) Please send your comments!

Mortal computation is described in [22]. It is computation which cannot be separated from the physical computing device, as 'immortal' software which could run on another device.

---

[2] The wave hypothesis will not solve the hard problem of consciousness [21]. But there are hard problems of physics (why does matter exist? What is the most fundamental particle?) and of cosmology (why does the universe exist?). These questions will never be answered, and we do science in spite of them.

I suggest that the wave representation of 3-D space is an example of mortal computing, because its computation and information storage cannot be separated from the physical medium that holds the wave. It is morphic, in that the physical form of the wave is essential to its function. Although the main role of the wave is information storage, it has a computational capacity in retrieval - retrieving selectively by spatial displacement (using wave interference). The wave is an analogue computing device.

You can see a discussion of the wave, and of this assessment at https://www.youtube.com/live/zqOcywx40n8 .

## 8. Further Tests of the Wave Hypothesis

The next step is further research to confirm or refute the existence of the wave. Initial suggestions for research are made in a companion paper [3]. That paper has a summary table of possible investigations of the wave hypothesis:

| Sub-sect. | Investigation |
|---|---|
| 3.1 | Thalamus connectome tests |
| 3.2 | Insect central body connectome tests |
| 3.3 | Detailed thalamic neuroanatomy |
| 3.4 | Detailed neuroanatomy of the insect central body |
| 4.1 | Multiple Species comparisons |
| 4.2 | Single-celled animal cognition |
| 5.1 | Targeted genomic and proteomic searches |
| 6.1 | Neural computational models of spatial cognition |
| 6.2 | Wave-neural computational models of spatial cognition |
| 6.3 | Computational models of sense data steering and binding |
| 7.1 | Insect behavioral studies |
| 7.2 | Small animal behavioral studies |
| 8.1 | Data from phenomenal consciousness |
| 9.3 | Quantum coherence in biological matter |
| 9.4 | Nuclear and electron spin effects |
| 9.5 | Bose-Einstein Condensates |
| 9.6 | Sustaining the wave, and coupling neurons to the wave |
| 10 | Direct detection of the wave |

This gives an idea of the wide range of disciplines that can contribute. The biophysical issues are preliminary and open-ended; it is not yet possible to say which ideas hold promise.

## References

With a few exceptions, this list only includes papers which directly discuss the wave hypothesis. References to other work are in those papers.

[1] Worden R.P. (2024) Spatial Cognition: A Wave Hypothesis, http://arxiv.org/abs/2405.10112

[2] Worden R.P. (2024) Three-dimensional Spatial Cognition: Bees and Bats, http://arxiv.org/abs/2405.09413

[3] Worden, R. P. Further tests of the brain wave hypothesis, unpublished paper, to be posted on arXiv.

[4] Worden R.P. (2024) The Projective Wave Theory of Consciousness, http://arxiv.org/abs/2405.12071

[5] McFadden, J. (2002) Synchronous firing and its influence on the brain's magnetic field. Journal of Consciousness Studies, 9, 23-50

[6] Pinotsis, D.A., Fridman, G., and Miller, E.K. (2023) Cytoelectric Coupling: Electric fields sculpt neural activity and "tune" the brain's infrastructure. Progress in Neurobiology, https://doi.org/10.1016/j.pneurobio.2023.102465.

[7] Sprenger J & Hartmann S (2019) Bayesian Philosophy of Science, Oxford

[8] Friston K. (2010) The free-energy principle: a unified brain theory? Nature Reviews Neuroscience

[9] Parr T, Pezzulo G, and Friston F (2022) Active Inference: The Free Energy Principle in Mind , Brain and Behaviour, MIT Press, Cambridge, Mass

[10] Parr T, Sajid N, Da Costa L, Mirza M & Friston K (2021) Generative Models for Active Vision, Frontiers in Neurobotics, 15

[11] Wallis G. and Rolls E.T (1997) Invariant face and object recognition in the visual system, Progress in neurobiology, 51,167

[12] Rolls, E.T., and Deco, G. (2002). Computational Neuroscience of Vision. Oxford: Oxford University Press.

[13] Hawkins J, Ahmad S. and Cui Y (2017) A Theory of How Columns in the Neocortex Enable Learning the Structure of the World, Frontiers in Neuroscience, doi: 10.3389/fncir.2017.00081

[14] Riesenhuber M. and Poggio T. (1999) Are Cortical Models Really Bound by the "Binding Problem"? Neuron, Vol. 24, 87–93

[15] Worden, R. P. (2020) Is there a wave excitation in the thalamus?  arXiv:2006.03420

[16] Worden, R.P. (2010) Why does the Thalamus stay together?, unpublished paper on ResearchGate

[17] Worden, R.P. (2014) the Thalamic Reticular Nucleus: an anomaly; unpublished paper on ResearchGate

[18] Worden, R. P. (2020). An Aggregator model of spatial cognition. arXiv 2011.05853.

[19] Worden R.P, Bennett M and Neascu V (2021) The Thalamus as a Blackboard for Perception and Planning, Front. Behav. Neurosci., 01 March 2021, Sec. Motivation and Reward, https://doi.org/10.3389/fnbeh.2021.633872

[20] Marr, D (1982) Vision. New York, NY: Freeman

[21] Chalmers, D. J. (1996). The conscious mind: in search of a fundamental theory. New York, Oxford University Press

[22] Ororbia A. and Friston K (2023) Mortal Computation: a Foundation for Biomimetic Intelligence, https://arxiv.org/abs/2311.09589